\documentclass{appolb}
\usepackage{epsfig}
\usepackage{axodraw}
\usepackage{url}

\begin{document}

\newcommand{\Rp}{${R_p \!\!\!\!\!\! / \,\,\,\,\,}$}

\title{Speculations on Isolated Lepton Events at HERA%
\thanks{Presented by J. Kalinowski  at the XXXI International Conference of Theoretical
Physics, "Matter to the Deepest", Ustro\'n, Poland, September 5–-11, 2007.}%
}
\author{S.Y. Choi$^{1,2}$, J. Kalinowski$^{3}$,
   H.-U. Martyn$^{1,4}$, R. R\"{u}ckl$^{5}$ and  H.~Spiesberger$^{6}$
\address{$^1$ Deutsches Elektronen-Synchrotron DESY, Hamburg, Germany\\
     $^2$ Physics Dept.\ and RIPC, Chonbuk National University, Jeonju, Korea\\
     $^3$ Institute of Theoretical Physics, University of Warsaw, Poland
          \\
     $^4$ I. Physikalisches Institut, RWTH Aachen, Germany\\
     $^5$ Institut f\"{u}r Theoretische Physik, Universit\"{a}t W\"{u}rzburg, 
           Germany\\ 
     $^6$ Institut f\"{u}r Physik, Johannes-Gutenberg-Universit\"{a}t Mainz,  Germany
}
}
\maketitle
\begin{abstract}
Speculations on mechanisms which might be responsible for events with an isolated high $p_T$ lepton, a hadron jet and missing energy,
as observed in the H1 experiment at HERA, are discussed.
\end{abstract}

\section{Introduction}
The first event with an isolated high $p_T$ lepton, a hadron jet and missing energy
\begin{eqnarray}
e^+ p\ \ \to\ \ e^+/\mu^+\, +\, \mbox{jet}\, +\, p_T^{miss}
\label{eq:event}
\end{eqnarray}
 observed at HERA by the H1 collaboration was announced more than 10 years ago \cite{Ahmed:1994mx}. An excess of HERA I events above {\it a priori} expectations (not confirmed by ZEUS, where the analysis was performed in a more restricted phase space, see~\cite{R1}) has given rise to many speculations on physics beyond the Standard Model. These events continued to appear in runs at HERA II. Both H1 and ZEUS have recently performed an analysis of the electron and muon channels (H1 also of the tau channel)
on their respective complete HERA I+II data sets, which correspond to approximately 0.5 fb$^{-1}$
per experiment \cite{R2,R3}. A total of 59 events are observed in the H1 data, compared to a SM
expectation of 58.9 $\pm$ 8.2. When the  hadronic transverse momentum cut $p^X_T > 25$ GeV is imposed, 24 events remain compared to 15.8 $\pm$ 2.5 expected, with 21 events observed in the $e^+p$ data and 8.9 $\pm$ 1.5 expected.
The observed data excess in the HERA I $e^+$p data thus remains at the
3$\sigma$ level for the complete H1  data set. In the ZEUS analysis of the complete HERA I+II
data, 41 data events are observed in agreement with 48.3 $\pm$ 6.8 expected. Unlike in
the H1 analysis, agreement between data and SM predictions is also observed in the high $p^X_T$ region, where
11 events are seen in the $e^\pm p$ data compared to 13.1 $\pm$ 1.8 expected.

\section{Theoretical Speculations }

A potential interpretation of these events  is
based on  $R$-parity violating supersymmetry (for a recent review of $R$-parity violating SUSY
see Ref.~\cite{R4}). Most speculations assume the formation of fairly light
stop particles in $e^+d$ fusion, via the $R$-parity violating
interaction $\lambda'_{131}L_1Q_3D^c_1$ in the superpotential, with stops
decaying to a $b$-jet and a chargino. (Scenarios with a charm squark produced in  $e^+d$
  fusion via the $\lambda'_{121}$ coupling and decays to a $s$-jet
  and a chargino can also be considered, although, due to possible strong mixing in the stop sector,
  the lightest stop can be expected considerably lighter than the charm squark.) The models differ in the structure of subsequent chargino decay chains.

\noindent (i) Charginos decay to a $W$-boson and a neutralino $\tilde{\chi}^0_1$,
or to a lepton and a sneutrino.   Subsequent
neutralino/sneutrino decays generate final
states more complex than in (\ref{eq:event}), unless the neutralino
or the sneutrino is assumed to be metastable,
decaying outside the detector. Such an interpretation
is very unlikely, however, since
the lifetime of $\tilde{\chi}^0_1$  is bounded from above by
the $\tilde{\chi}^0_1 \to b \tilde{b} \to b d \nu$ channel mediated
by  a virtual $\tilde{b}$, and
the sneutrino may decay ``back'' through the channel $\tilde{\nu}_\tau\to
\tau\tilde{\chi}^+_1 \to \tau b \tilde{t} \to \tau b e^+d$ involving virtual
intermediate $\tilde{\chi}^+_1$ and $\tilde{t}$ states.
The estimated decay widths of the two modes, $\Gamma(\tilde{\chi}^0_1)
\sim 1 $ eV and $\Gamma( \tilde{\nu}_\tau) \sim 10^{-3}$ eV, for
$\lambda'_{131}\sim 5\times 10^{-2}$ and SUSY masses ${\cal O}$(100) GeV needed to explain the production rate,
suggest lifetimes of order $\sim 10^{-15}$ sec and $\sim 10^{-12}$ sec,
respectively, much too short for the particles to escape the detector before
decaying \cite{Choi:2006ms}.

\noindent (ii) An alternative interpretation is based on  $R$-parity
violation (\Rp) in both lepton-quark and lepton-lepton interactions in which  the
chargino decays to a charged slepton
and a neutrino, followed by the subsequent slepton \Rp decay to a
lepton$+$neutrino pair, giving the characteristic final state of (\ref{eq:event}), cf. Ref.~\cite{R5}.
The production of the intermediate stop particle is governed by the
\Rp term $\lambda'\,LQD^c$ in the superpotential, while the slepton
decays to a lepton $+$ neutrino pair by $\lambda\,LLE^c$ \cite{Choi:2006ms}.

\section{An \boldmath{$R$}-Parity Violating Stop Scenario}

The interpretation (ii) requires: (a)
the coupling $\lambda'$ sufficiently large, a few times $10^{-2}$ for
$\lambda'_{131}$, to guarantee the necessary stop production rate; (b) the intermediate particles
generated on-shell and the coupling $\lambda$ sufficiently large since  otherwise the
rates of events (\ref{eq:event}) fall dramatically; (c) the light chargino and the lightest neutralino
higgsino-like to suppress $\tilde{\chi}^+_1 \to
W^+\tilde{\chi}^0_1$ decays which
would not generate the desired
final states; (d) the Higgs mixing parameter $\tan\beta$ moderate
to allow comparable charged slepton and sneutrino decays of the chargino.\\[-5mm]
\begin{table}[htb]
\caption{\label{tab:tab1} {\it Definition of the reference point $\Re$.
  }}
{
\begin{center}
\begin{tabular}{|l||l|}
\hline
\ \ \ \ $\Re$: Parameters\ \ \ \   &\ \ \ \  Values     \\
      \hline\hline
\ \ \ \    elw gaugino masses     &\ \ \ \  $M_2=2M_1=1.5$ TeV \\
\ \ \ \    higgsino mass          &\ \ \ \  $\mu=160$ GeV  \\
\ \ \ \    Higgs mixing           &\ \ \ \  $\tan\beta =1.5$ \\
       \hline
\ \ \ \    scalar lepton masses   &\ \ \ \  $M_L = M_E=130$ GeV \\
\ \ \ \    scalar quark masses    &\ \ \ \  $M_Q = M_U = M_D = 420$ GeV\\
\ \ \ \    trilinear $A$ coupling &\ \ \ \  $A_t = 840$ GeV \\
       \hline\hline
\ \ \ \    $\lambda', \lambda$ couplings
                  &\ \ \ \ $\lambda'_{131}=\lambda_{322}=\lambda_{321}=5\times 10^{-2}$ \\
\ \ \ \    { }    &\ \ \ \ other $\lambda',\lambda$ very small\\
       \hline
\end{tabular}
\end{center}
}
\end{table}\\[-5mm]
The reference point $\Re$ defined in Table 1 \cite{Choi:2006ms} does not appear in
conflict with experiment. It is chosen to
develop constraints on the  \Rp interpretation of experimental data but
it should not be mis-interpreted as an outstanding candidate for explaining
existing data. Masses and mixings generated by this reference point  are compatible with the bounds on masses and mixings
from LEP, Tevatron and HERA. Strictly within the
MSSM, the parameters would lead to too low a mass of the lightest Higgs
boson; however, this mass can be raised beyond the LEP limit in
extended theories without affecting the mass and mixing
parameters relevant for our discussion.

\begin{figure}[htbp]
\begin{picture}(100,170)(-50,60)
\put(50,185){$e^+d \longrightarrow \tilde{t}\longrightarrow\tilde{\chi}_1^+ + b$}
\ArrowLine(125,176)(56,146)
\ArrowLine(125,176)(194,146)
\put(38,131){$\tilde{\ell}^+_L + \nu$}
\put(180,131){$\ell^+ + \tilde{\nu}$}
\ArrowLine(42,120)(-1,90)
\ArrowLine(42,120)(42,90)
\ArrowLine(42,120)(85,90)
\ArrowLine(209,120)(166,90)
\ArrowLine(209,120)(209,90)
\ArrowLine(209,120)(252,90)
\put(-15,80){$e^+\nu_\mu$}
\put(-15,65){$e^+\nu_\tau$}
\put(35,80){$\mu^+\nu_\mu$}
\put(35,65){$\mu^+\nu_\tau$}
\put(83,80){$t\bar{d}$}
\put(148,80){$e^-\mu^+$}
\put(199,80){$\mu^-\mu^+$}
\put(250,80){$d\bar{b}$}
\put(148,65){$e^-\tau^+$}
\put(199,65){$\mu^-\tau^+$}
\end{picture}
\caption{\label{fig:tree} \it Mixed $R$-parity conserving and $R$-parity
  violating decays of the lighter top squark $\tilde{t}_1$ which give
  rise to multi-lepton and jet final states with missing
  transverse momentum (left cascade) due to escaping neutrinos. }
\end{figure}
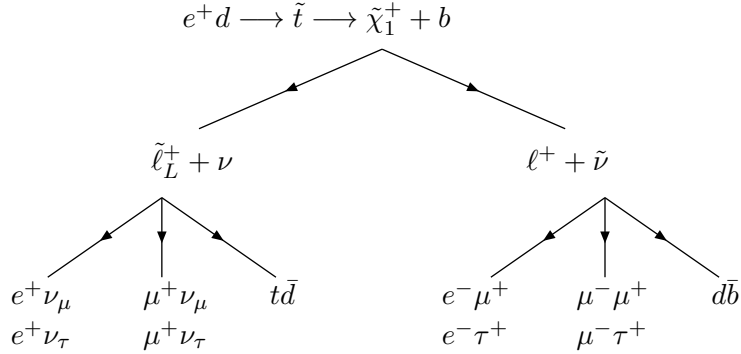

Since
the \Rp couplings of the superfields
generate several interactions in the (s)quark and (s)lepton sectors of different charges and spieces, a large variety of experimental signatures can be expected. Events including multi-lepton final states and $\tau^+$ leptons should be observed,  see Fig.\ref{fig:tree}, with
no overwhelming preference for a particular channel but with  large uncertainties
due to the
choice of mass, mixing and $\lambda', \lambda$ parameters.

\section{Constraints on the  SUSY \Rp Interpretation}
Many
consistency conditions for the proposed SUSY scenario
can be derived from the kinematic properties of leptons and jets in the final state.
Using the $\tilde{t}_1$ 4-momentum from the on-shell $\tilde{t}_1$ production
in $e^+d$ collisions
and the measured  $b$-jet energy
$E_b$ and longitudinal $z$-momentum $p_b^z$ along the proton
direction, the $\tilde{\chi}^+_1$ mass condition
$m^2_{\tilde{\chi}^+_1}=(p_{\tilde{t}_1}-p_b)^2$ can be cast in the
form
\begin{eqnarray}
m^2_{\tilde{\chi}^\pm_1} \ \ = \ \
m^2_{\tilde{t}_1}[1 - (E_b-p_b^z)/2E_e] -2 E_e(E_b+p^z_b) \label{eq:onshell}
\end{eqnarray}
With the {\it a priori} unknown stop and chargino masses,
each event with measured values of
$b$-jet energy and longitudinal momentum defines,
according to Eq.$\,$(\ref{eq:onshell}), a line in the
(mass)$^2$ plane, the coordinates labeled by $(m^2_x,m^2_y)$.
If the
considered \Rp scenario is the correct interpretation of the data, lines corresponding to the signal events must cross at the single point
corresponding to the true values of stop and chargino masses
$(m^2_x,m^2_y)=(m^2_{\tilde{t}_1},m^2_{\tilde{\chi}^\pm_1})$, and lie 
within the double cone formed by the line with slope = 1 and
cutting the $m^2_y$-axis at $-(m^2_{\tilde{t}_1}-
m^2_{\tilde{\chi}^\pm_1})$, and the line crossing the origin with
slope = $m^2_{\tilde{\chi}^\pm_1}/m^2_{\tilde{t}_1}$, both intercept and
slope given by the true mass values. The opening angle of the
cone increases apparently with the $\tilde{t}_1 - \tilde{\chi}^\pm_1$
mass gap.

\begin{figure}[htb!]
\begin{center}
\includegraphics*[height=5cm,width=10cm,angle=0]{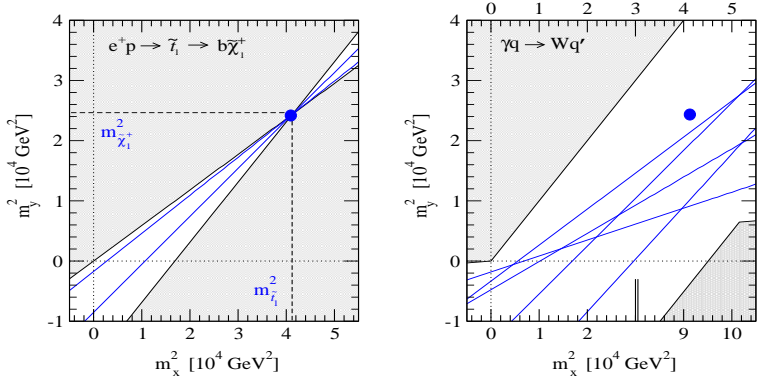}
\end{center}
\vskip -0.3cm
\caption{\it Left:
 All \Rp signal lines  must lie within the double cone and
 cross at the true mass values,
 $m^2_x \to m^2_{\tilde{t}_1}$ and $m^2_y \to m^2_{\tilde{\chi}^\pm_1}$.
 Right: Lines corresponding to background events from $W$-photoproduction
 $\gamma q\to Wq' $. The kinematically excluded regions are shaded (the lower $x$-axis is cut to accommodate the
 forbidden wedge on the right).}
\label{fig:xcross}
\end{figure}

On the other hand, lines corresponding to background events do not
cross at a single point but rather form an irregular mesh over the
$(m^2_x,m^2_y)$ plane. This is shown in Fig.$\,$\ref{fig:xcross} for a
few examples of signal events (left), and for background (right) for which the jet parameters are identified with values
derived from $W$-photoproduction $\gamma q\to Wq'$.

The on-shell requirements provide many more consistency conditions.  For example, in the decay chain ending with $t\bar d$ the events must cluster at the invariant mass
$M[jt]= m_{\tilde{e}_L}$, while in the decay chain ending with  $d\bar b$ they must cluster at the triple point $M[jj]= m_{\tilde{\nu}_e},\,
M[e^+ jj] = m_{\tilde{\chi}^+_1},\,
M[e^+ jjj]= m_{\tilde{t}_1}$ ($j$ stands for a non-$t$ jet). More constraints can be derived for final states with $\tau$ since the $\nu_\tau$ 3-momentum can be reconstructed fully in
single $\tau$ events.

\section{Conclusions}
 Independent of the specific reference point, two generic implications
have emerged from the study:\\
-- Kinematical constraints relate the observed jet and lepton
      energies and momenta with masses of stop, chargino and sneutrino, and
      clusters of invariant  masses must be observed experimentally;\\
-- $R$-parity violating couplings connect leptons and sleptons of different
      charges and species, implying that also events including multi-lepton
      final states and $\tau^+$ leptons should be observed.

\subsubsection*{Acknowledgments}
Work supported by the Polish Ministry of Science and Higher Education
Grant 1 P03B 108 30,  the Korea Research Foundation Grant funded by the 
Korean Government (MOEHRD, Basic Research Promotion Fund) (KRF-2007-521-C00065), 
and the EU  Network MRTN-CT-2006-035505 
"Tools and Precision Calculations for Physics Discoveries at Colliders".

\end{document}